\documentclass[reprint, amsmath,amssymb, aps, prb]{revtex4-2}

\usepackage{graphicx}
\usepackage[colorlinks=true, allcolors=blue]{hyperref}
\graphicspath{{Images/}}

\newcommand{\be}{\begin{equation}}
\newcommand{\bea}{\begin{eqnarray}}
\newcommand{\ee}{\end{equation}}
\newcommand{\eea}{\end{eqnarray}}

\def\L{\mbox{$L$}}

\def\T{\mbox{{\bf $T$}}}

\def\I{\mbox{$\bf{I}$}}
\def\balpha{\mbox{$\boldsymbol{\alpha}$}}

\newcommand{\eg}{{\it e.g.}}
\newcommand{\ie}{{\it i.e.}}

\begin{document}

\title{Machine Learning Classification Informed by a Functional Biophysical System}

\author{Jason A. Platt}
 \thanks{Corresponding Author: jplatt@ucsd.edu}
\author{Anna Miller}
\author{Lawson Fuller}
\affiliation{Department of Physics,\\
University of California San Diego\\
9500 Gilman Drive\\
La Jolla, CA 92093\\
}

\author{Henry D. I. Abarbanel}
\affiliation{Department of Physics,\\
and\\
Marine Physical Laboratory,\\
Scripps Institution of Oceanography,\\
University of California San Diego\\
9500 Gilman Drive\\
La Jolla, CA 92093\\}

\date{\today}

\begin{abstract}
We present a novel machine learning architecture for classification suggested by experiments on olfactory systems. The network separates input stimuli, represented as spatially distinct currents, via winnerless competition---a process based on the intrinsic sequential dynamics of the neural system---then uses a support vector machine (SVM) to provide precision to the space-time separation of the output. The combined network uses biophysical models of neurons and shows high discrimination among inputs and robustness to noise. While using the SVM alone does not permit determination of the components of mixtures of classified inputs, the combined network is able to tell the precise concentrations of the constituent parts.
\end{abstract}

%\keywords{Suggested keywords}%Use showkeys class option if keyword
                              %display desired
\maketitle
\section{Introduction}
Supervised machine learning (ML) networks are frequently tasked with the recognition and classification of objects or mixtures of objects.  Doing so requires the establishment of i) a network architecture and ii) a learning algorithm.   Generally, the network consists of nonlinear active units (`neurons'), arranged in layers, with rules governing how neurons in one internal layer are influenced by the activity of the neurons in another. Matching inputs and associated outputs---constrained by layer-to-layer rules---provide a guide for the success of the network's selection capability.  

Goodfellow and coauthors \cite{goodfellow16} make the case that `deep' networks of this sort, with many layers, move the focus from complex brain structures to `representations' of and within the data: ``{\bf Deep learning} solves the central problem in representation learning by introducing representations that are expressed in terms of other, simpler representations.''  Their Fig 1.4 ~\cite{goodfellow16}  gives an overview of the development of ML algorithms showing their interpretation of the varying levels of abstraction and representation in artificial intelligence.  While `large enough' networks of this sort have shown substantial success in prediction~\cite{gulshan16}~\cite{goodfellow16}, the conventional architecture and backpropagation learning rule is different from how neurobiological networks perform the same tasks. 

In this paper we take a step back and examine a small, biologically inspired network as the inspiration for a novel ML classifier. ``Small'' functional biological networks---units of  $\sim 10^3 - 10^6$ neurons that have evolved to perform a specific task, such as odor classification or song production---have been the object of intense experimental study over the last few decades \eg, \cite{Margoliash95}. Often, many details about the biophysical operations of these smaller networks are known from laboratory experiments; therefore, aspects of the neural circuitry can serve as a guide for generating an ML network architecture and learning rule.  In this paper we analyze some of the capabilities of one of these small functional networks, namely, the insect olfactory system. This biological system functions to separate chemical constituents, in both time and network space, of an `odor' represented as distinct stimulating currents projected forward from sensors in an antenna.  

Huerta and Nowotny~\cite{huerta09} were the first, to our knowledge,  to suggest that the neural circuits of the insect olfactory system might serve as a model for classification tasks in ML. Recognizing that biological olfactory networks rapidly and accurately identify chemical constituents in odors, they argued that an abstraction of these networks could perform the same classification task on common data sets in the ML literature~\cite{Huerta2013}.  

We provide a concrete instantiation of these suggestions and demonstrate how they may be effectively used in an ML context. Our aim is not to improve on the excellent modeling and numerical analyses of the network stages of the biological olfactory system of insects or mammals---rather, it is to adopt the functional aspects of those stages into a usable model framework. Furthermore, we investigate how these elements exhibit two important operational aspects: (1) robustness against noise in the stimulating currents after the network has been trained, and (2) the manner in which the network output permits, in a systematic manner, accurate identification of the amplitudes of components of mixtures of learned sets of `odors' (currents).

The extensively studied insect olfactory network is employed as a guide---informing the construction of the ML device.  Numerous studies both experimental~\cite{Fernandez2009, Laurent1996, Laurent1999} and computational~\cite{ahn2010, assisi_adaptive_2007, sanda_classification_2016, bazhenov_fast_2005, bazhenov01a, bazhenov01b, Bazhenov2015,huerta09} have characterized the insect olfactory system in terms of its ability to classify odors and mixtures of odors.  With some certainty then, one can proceed with both a biophysical and mathematical understanding of how this network operates. This approach to the classification task incorporates the design of functional networks operating under well understood biophysical principles. This insight into the `black box' of ML enables manipulation and control of the classification process, and allows alteration of the timing and speed of the network dynamics.

\section{Olfaction}
\begin{figure}[h!tbp]
  \centering
  \includegraphics[width=0.5\textwidth]{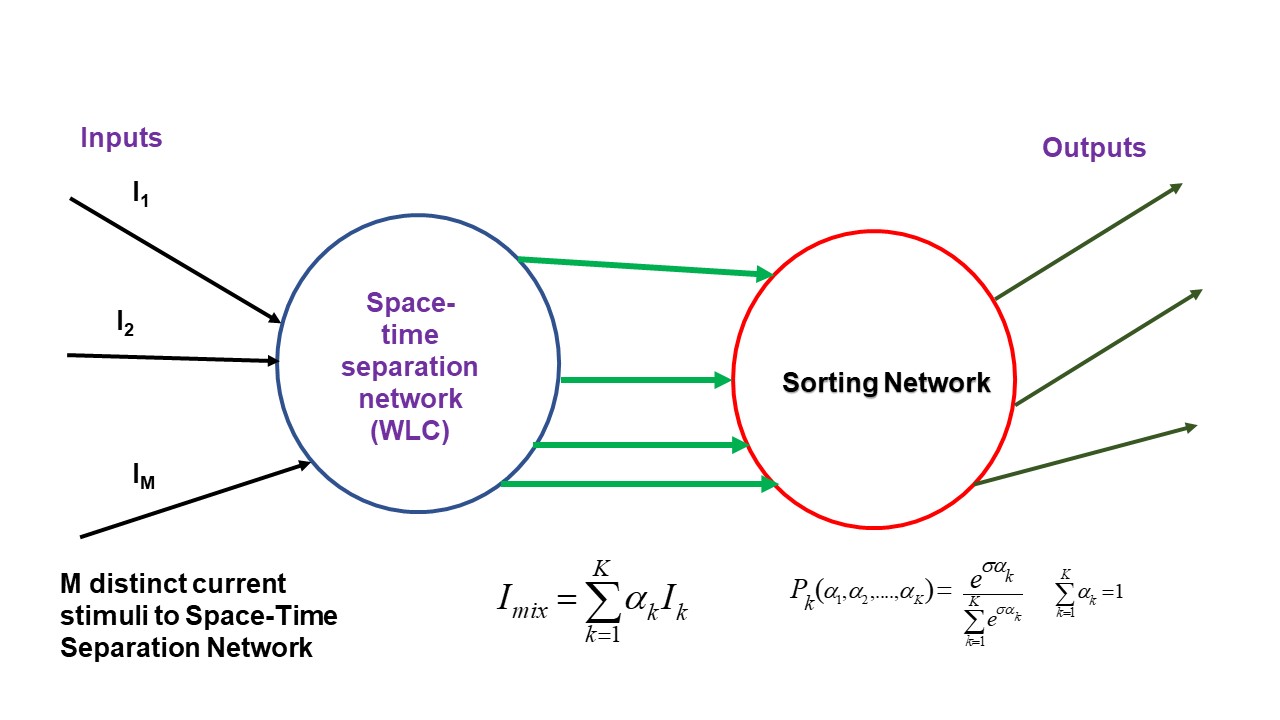}
  \includegraphics[width=0.5\textwidth]{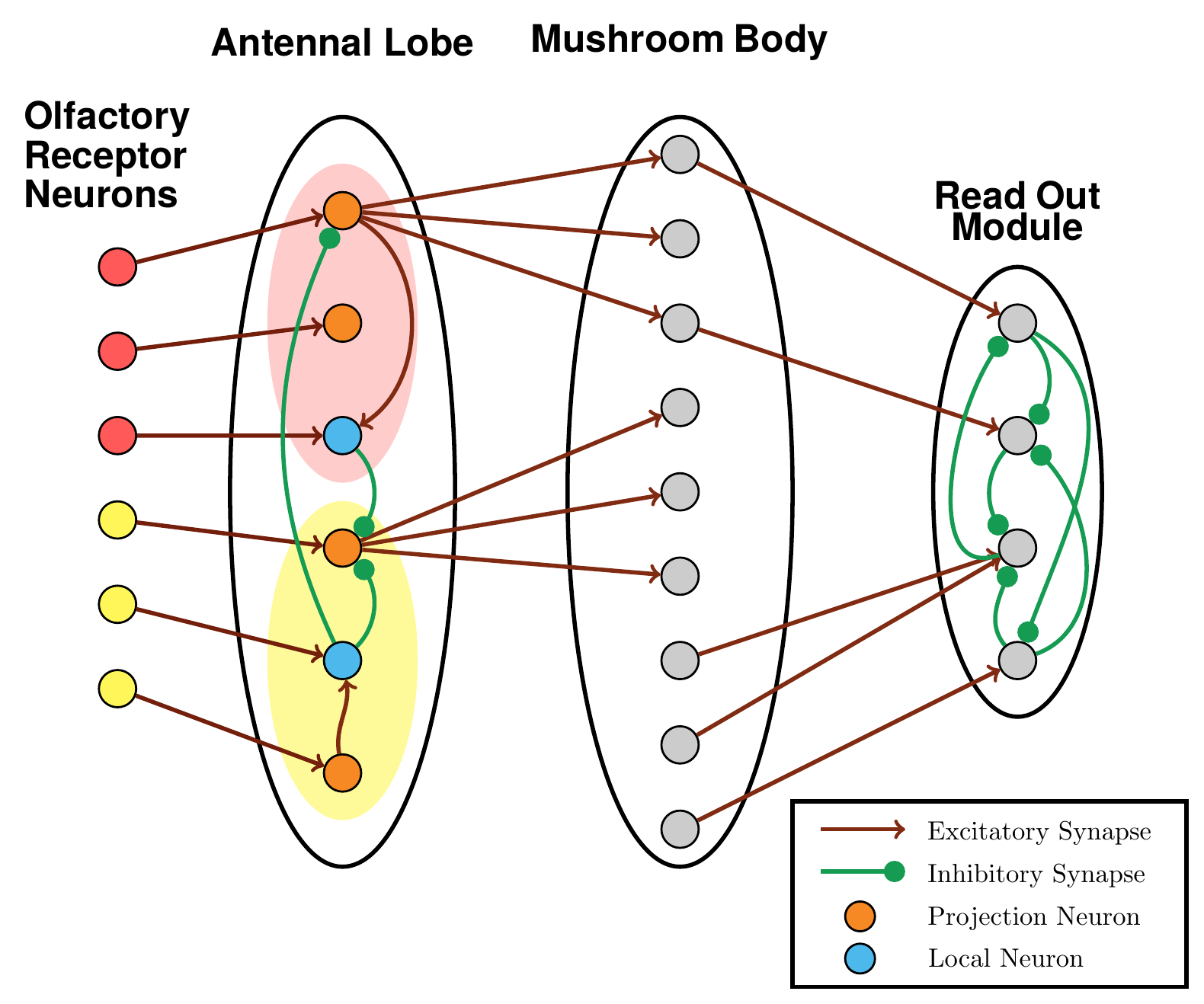}
  \caption{\textbf{(Top Panel)} Structure of the olfactory inspired classification network presented in this paper.  Distinct input currents $I$---Eq. \eqref{eq: input}---are presented to the WLC network that has N $\sim$ $\mathcal{O}(1000)$ active units. Each stimulus defines an ``initial direction'' and the trajectories they stimulate are shown in Fig. (\ref{fig: pca_HH}) when M = 5.  The WLC feeds into a sorting network---here an support vector machine (SVM)---which classifies the separated inputs or mixtures of those inputs. \textbf{(Bottom Panel)} The biological neural circuits that perform the identification and classification of components in odors. The initial stage is composed of olfactory receptor neurons neurons (ORN) that are activated when a particular chemical component attaches to these neurons. These ORNs produce electrical signals directed to the second stage called the antennal lobe (AL).  In application to other classification tasks, the sensors producing the currents could be acoustic, optical, or other modalities.  Within the AL are excitatory projection neurons (PNs) and inhibitory interneurons (LNs). In locusts there are approximately 850 PNs and 300 LNs.  The PNs carry AL activity forward to the next stage of olfactory recognition called the mushroom body (MB), which is suggested to act as a support vector machine in the biological olfactory network \cite{Huerta2013} . In locusts there are estimated to be 50,000 cells in the MB~\cite{gilles01}.}
  \label{fig: bio_net}
\end{figure}

The simplified structure of the ``front end'' of the insect olfactory system is sketched in Fig (\ref{fig: bio_net}). This sketch does not go into detail; however, see~\cite{gilles01}.  Our construction for ML purposes does not strive to reproduce details of observed biological networks, only to abstract and utilize their essential functionality.  This paper focuses on the functional roles of the antennal lobe (AL) and the mushroom body (MB).  

%fig1

When stimulated by distinct current inputs from a sensory input network, the biological AL produces trajectories in the network phase space following distinct heteroclinic sequences among unstable regions. These trajectories move from unstable regions to other unstable regions, and continue to do so as long as the stimulus persists. When the stimulus ends, the trajectory retreats to a stable fixed point region where it responds to environmental noise~\cite{Laurent1996,Laurent1999,LaurentWehr1996}. These observed properties led to a suggestion of an AL network structure~\cite{rabinovich2000,rabinovich2001,rabinovich2003} called {\bf winnerless competition networks} (WLC).

The idea of the AL as a WLC network was examined in experiments on locust olfactory networks by Mazor and Laurent~\cite{mazor05}. They found that the AL responses to stimulating odors of varying duration were described by: (1) an ``on-transient'', when the stimulus is first received, (2) an ``off-transient'' as the stimulus recedes and the neural activity returns to its stable base state, and (3) movement around a `fixed point' or stable region in AL neuron phase space. They noted that ``optimal stimulus separation occurred during the transients\ldots'', suggesting that the biological AL acts as a WLC network with added longer time scale activity and odor specific dynamics. 

Once the WLC structure is established, the precise trajectory of the network response in network state space and time is determined by the specific stimulus.  Each distinct stimulus defines a specific direction in the high dimensional space of the WLC network.  As the network state space is composed of multiple regions of nonlinear, unstable behavior, the phase space trajectory is seen to be quite sensitive to the selected current stimulus. This sensitivity suggests that a WLC network could distinguish among many `nearby' stimuli. Indeed, there is an estimation of the capacity of $e(N-1)!$ for a WLC network of N neurons~\cite{rabinovich2001}.

Olfaction also employs a second stage of classification network, possibly because it allows the specification and separation of phase space regions in a more precise manner than utilizing the AL alone. This is represented here by a support vector machine (SVM)~\cite{vapnik92, vapnik95} to produce a classifier operating on the AL projection neuron outputs. Huerta~\cite{Huerta2013} also suggests that the second stage network might act as an SVM in insect olfaction.

\begin{figure}[h!tbp]
  \centering
  \includegraphics[width=0.5\textwidth]{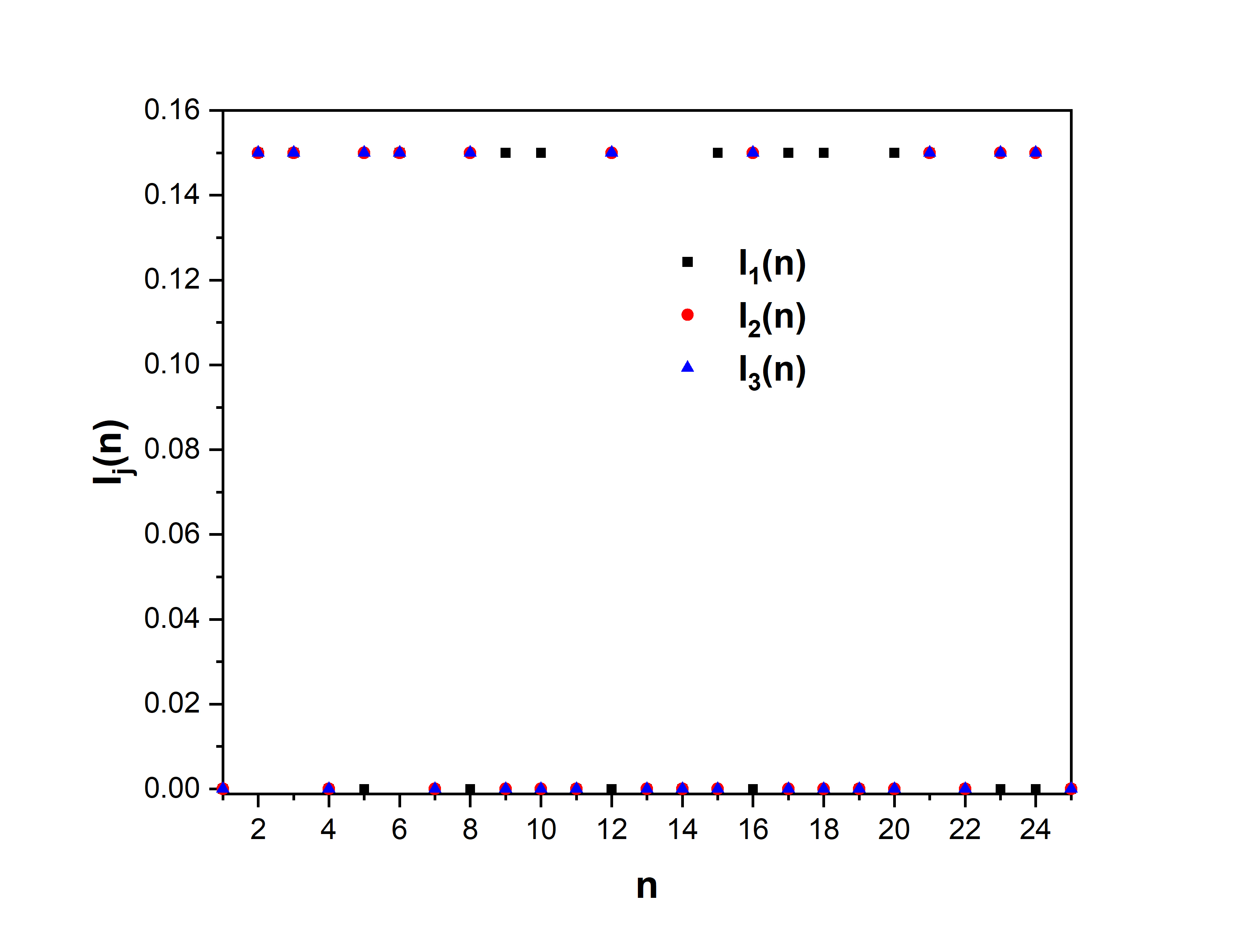}
  \caption{Examples of three distinct currents that are injected into the 1000 dimensional WLC network to stimulate activity in network space as defined in Eq. (\ref{eq: input}).  The x-axis defines $n$, the particular input neuron, with the y-axis showing that the input is either 0, or 0.15 $nA$. Only a segment of this current is shown; it is a 1000 dimensional vector with 1/3 of its entries populated at random with amplitude $I_0 = 0.15 nA$. There are $\binom{1000}{300}$ $\approx 5.42 \times$10$^{263}$ of these currents.}
  \label{fig: three_distinct_currents}
\end{figure}

\begin{figure}[h!tbp]
  \centering
  \includegraphics[width = 0.5\textwidth]{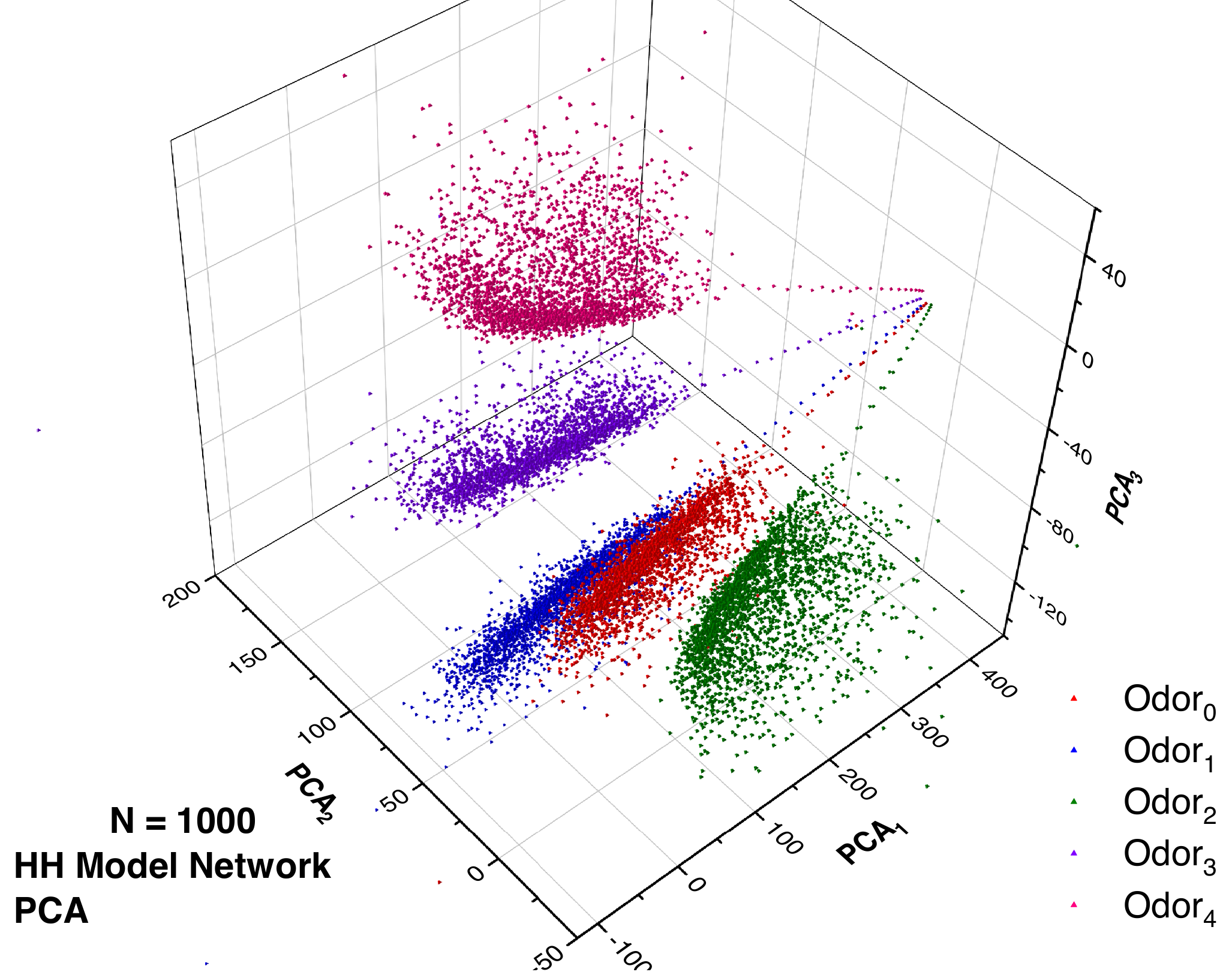}
  \caption{Five distinct input currents Eq. \eqref{eq: input} were presented to the 1000 model WLC network built with Hodgkin-Huxley (here) or FHN biophysical neurons~\cite{jwu,willshaw,biocyb1, FitzHugh1961, Nagumo1962} connected by inhibitory synaptic processes.  A PCA projection was performed on a concatenation of the five distinct, 1000 dimensional voltage signals produced by these stimuli. The projection displayed is into the three dimensional space spanned by the PCA eigenvectors with the largest singular values. In this very low dimensional space the separation of the input stimuli in space (neuron number) and time is clear. Before the stimulation is turned on, all signals are at rest. When stimulation begins they move rapidly to different regions of PCA projected state space and remain there while the stimulation persists.}
  \label{fig: pca_HH}
\end{figure}

We start with a reminder how WLC networks can be built, and explore the properties and performance of two such networks.  Each network has the same basic WLC architecture, but differ in that one uses FitzHugh Nagumo neurons (FHN)~\cite{FitzHugh1961, Nagumo1962}---Eq. \eqref{eq: FHN}---at its nodes, while the other utilizes simple Hodgkin-Huxley (HH)~\cite{jwu,willshaw} neurons with Na, K, and Leak channels. Results from both WLC networks are discussed; they are qualitatively similar.  The WLC network---representing an AL---contains $N \sim 1000$ neurons, with its performance determined by its ability to classify distinct `odors' represented by input currents. Once this WLC/AL is constructed, it remains fixed in performing its task of signal separation; it does not require training.

\section{Building a WLC Network for Classification} \label{sec: WLC}
\subsection{Construction}
WLC networks were constructed using the Brian development package~\cite{Brian}, and selecting FHN or HH neurons at the nodes of the network. Inhibitory synapses connect each presynaptic/postsynaptic pair. The strength of this inhibition is an important parameter in the WLC network.  The network examined here has N = 1000 neurons, approximately the complement in the locust AL~\cite{gilles01}, each connected randomly with probability = 50 \%.

The FHN neurons satisfy the dynamical equations
\bea
    &&\frac{dV_{post}(t)}{dt} = \frac{1}{\tau_1}(V_{post}(t) - \frac{V_{post}(t)^3}{3} - w(t) + \\
    &&I_{\rm inj}(t)- z(t) (V_{post}(t) - \nu) + 0.35 mV)\nonumber \\
    &&\frac{dw(t)}{dt} = V_{post}(t) - b w (t)+ a \nonumber \\
    &&\frac{dz(t)}{dt} = (I_{\rm syn}(t) - z(t))/\tau_2,
    \label{eq: FHN}
\eea
for each presynaptic/postsynaptic pair in the network.
The parameter values are $\nu = -1.5 mV$, $a = 0.7$, $b = 0.8$, $\tau_1 = 0.08 ms$, and $\tau_2 = 3.1 ms$.  $I_{\rm syn}(t) = g_{\rm nt}\sum_{V_{\rm pre}} \theta(V_{pre}(t))$. $\theta(u)$ is the Heaviside step function.  $g_{\rm nt} = 0.1$ is the critical parameter in the WLC network; it must be scaled with the size of the network and the number of connections between neurons.

\subsection{Input/Output}
Our `functional classifier' is a WLC network whose output is comprised of N voltage time series $V_n(t);\; n = 1, 2,\ldots,N$ for each distinct stimulus. In the insect the original stimulus is chemical; in other settings the physical origin of the stimulus may differ. Optical, acoustic, or other sensors producing distinct currents may be used for a variety of applications of the classifier structure we are building here.

What we call the M ``baseline'' current stimuli are $\I_m \; m=1,2,\ldots,M$ currents in N-dimensional neuron space is given by a distinct input vector which is zero for $t \le t_0$, zero for $t \ge t_{final}$, and DC in between:
\be 
\I_m(t) = I_0 \biggl [(X_1,X_2,\ldots,X_N)_m+ \eta \, \mathcal{U}(-1,1)]\biggr].
\label{eq: input}
\ee
$I_0$ is a constant; $\mathcal{U}(-1,1)$ is a uniform distribution with a range $[-1,1]$; $\eta \ge 0$.  The input base current vectors $X_n$ are selected by drawing the components of each $X_n: \{X_1,X_2,\ldots,X_N\}$ from a Bernoulli distribution such that $\sim$ 1/3 of the values are set to $1$ with the rest are set to $0$.  This gives $\sim \binom{N}{N/3}$ distinct possible combinations.  Fig. (\ref{fig: three_distinct_currents}) shows a graphical representation of the input currents.  The currents in Eq.(\ref{eq: input}) also have added noise $I_0 \eta \, \mathcal{U}(-1,1)$ to all $N$ neurons. As the noise level $\eta$ in a baseline current increases, one can inquire how well the network still classifies the noisy stimulus as the $\eta = 0$ baseline current.   The signal to noise ratio is $1/\eta \sqrt{3}$---see Fig. (\ref{fig: class_acc}).  References to baseline currents specifically refer to Eq. \eqref{eq: input}.

In section \ref{sec: mix2} and \ref{sec: mix3} we present results for mixtures of these baseline currents.  In section \ref{sec: time_dep} we multiply a time dependent component to the baseline currents to test the network performance on temporally varying inputs.

\subsection{Low Dimensional Projections of the 1000 Neuron WLC Network: PCA}
Visualizations of the distinct phase space trajectories in $N \sim 1000$ dimensions is not a feasible task. For this purpose only, the neuron voltage outputs $V_n(t);\;n=1,2,\ldots,N$ can be projected into low dimensional spaces using Principal Component Analysis (PCA)~\cite{Press-Flannery-2007-NumRecipes}. Projection to three dimensions produces compelling evidence that the WLC network yields distinguishable trajectories for distinct odors. Fig.(\ref{fig: pca_HH}) shows that for five different baseline odors, the WLC separation in space and time is apparent in the three dimensions selected by PCA. More accurate separation of input signals is achieved by the SVM---the next stage of the classification circuitry---in $N$ dimensional space.

The three axes---onto which the network output voltages are projected---are chosen by building a time series of matrix dimension $M(T+1) \times N$ ($M$: number of input currents, $N$: number of neurons, $T$: number of time steps the stimulation is on) resulting from concatenating the $M$ distinct current time series. PCA is performed on this matrix, and these data are projected into a space associated with the three largest singular values. This protocol mixes the currents together and decorrelates their contribution to the overall set of classes.

The three dimensional plots are not quantitative but only `suggestive,' as there is no {\it a priori} reason the results of the network voltage output signals should be separated, identifiable, or classifiable in such low dimensions. The strategy is to use the WLC power of class separation followed by an SVM operating in high dimensions (here $N \sim 1000$) to perform the precise identification required. 

\subsection{Support Vector Machine}
The separation in phase space and time of the network activity for distinct baseline inputs, apparent in Fig.(\ref{fig: pca_HH}), lends itself to the idea of following the WLC network by an SVM to allow additional precision in classification. This idea is suggested by the operation of the mushroom body, which projects the activity of the AL to a much higher dimensional space where the response is quite sparse.  The operation of a linear SVM finds the optimal separating hyperplanes between sets of points through a convex optimization process.  Translating the results to an ML device is straightforward. As only the SVM is trained, the WLC is held fixed and not trained. Well developed and documented methods can be used to train the SVM using a ``one vs rest'' multiclass classification algorithm~\cite{convex,cs229,winstonsvm}.

\section{Results} \label{sec: results}

\subsection{Robustness to Noise}
Holding the architecture of the WLC network fixed, the combined ML network (WLC+SVM) is trained on $M$ distinct baseline currents $I_m(t);m=1,2,\ldots,M$; $\eta = 0$ Eq.\eqref{eq: input}. Each of these odors is presented once to the WLC+SVM for $t_f - t_0$ ms.  SVM classification results in $M$ separating hyperplanes defining the domains of each of the baseline currents.

\begin{figure}[h!tpb]
    \centering
    \includegraphics[width = 0.5\textwidth]{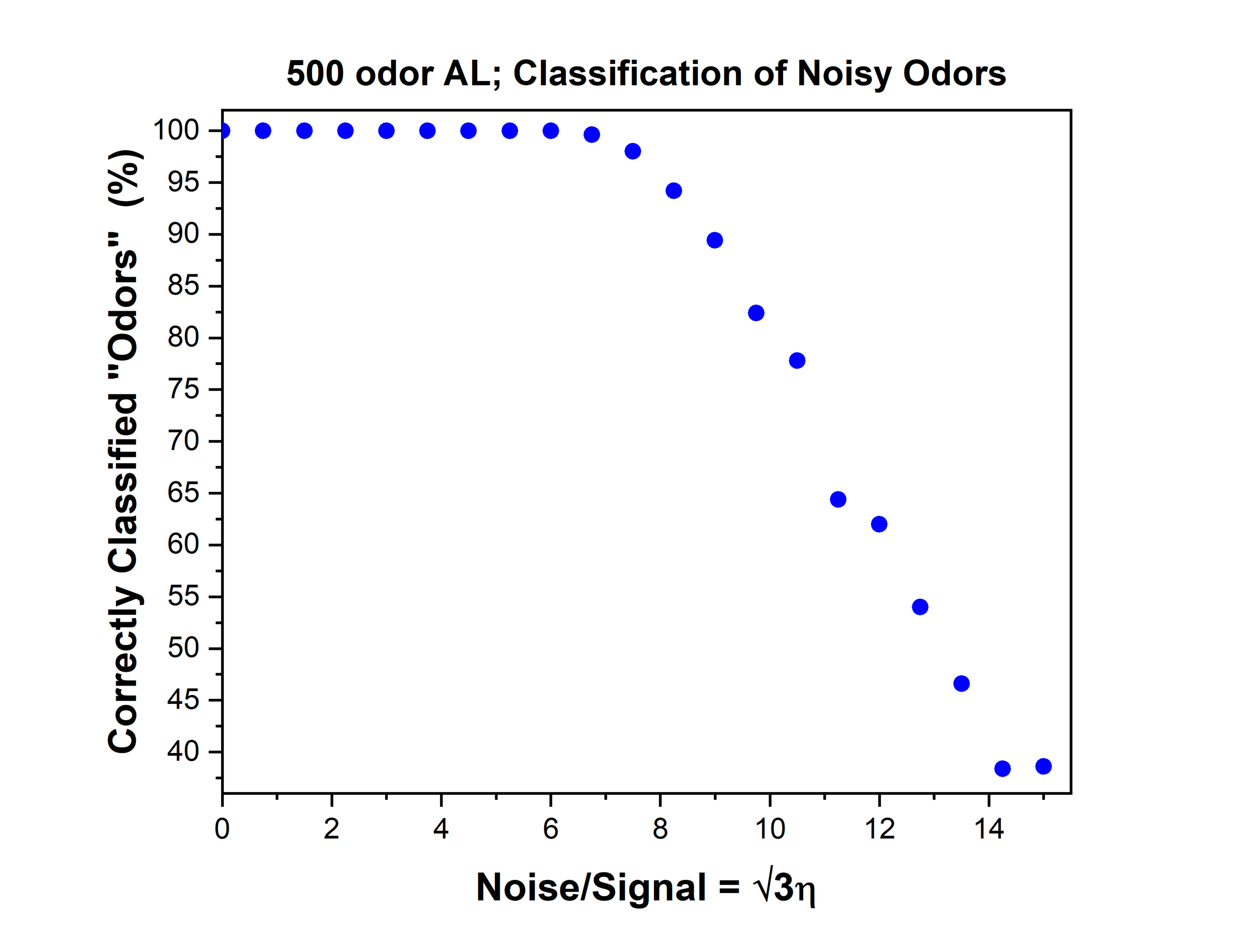}
    \caption{The classification accuracy of noisy currents presented to a WLC+SVM trained on 500 baseline ($\eta = 0$) currents---see Eq. \eqref{eq: input}. Multiple trials of each noisy ($\eta > 0$)  current were presented to the WLC+SVM network (with FHN neurons at the nodes) for $t_{final} - t_0 = 125\,$ ms. The current (odor) is labeled by taking the highest probability of Eq.\eqref{eq: prob}}
    \label{fig: class_acc}
\end{figure}

The robustness of the network is tested by measuring the accuracy of classification as a function of the number of classes and $\eta$---results shown in Fig.(\ref{fig: class_acc}). The network shows a high robustness to input noise, with accuracy slowly tailing off once the SNR reaches $\sim$ 0.16 (-16.9 dB). While we do not have a quantitative statement about the high robustness to additive noise, one may attribute this property to the large number of negative Lyapunov exponents in a WLC network that shape the trajectory as it moves from one unstable region to another~\cite{rabinovich2001}.

\subsection{Mixtures of Two Currents} \label{sec: mix2}
\begin{figure}[h!tpb]
    \centering
    \includegraphics[width = 0.5\textwidth]{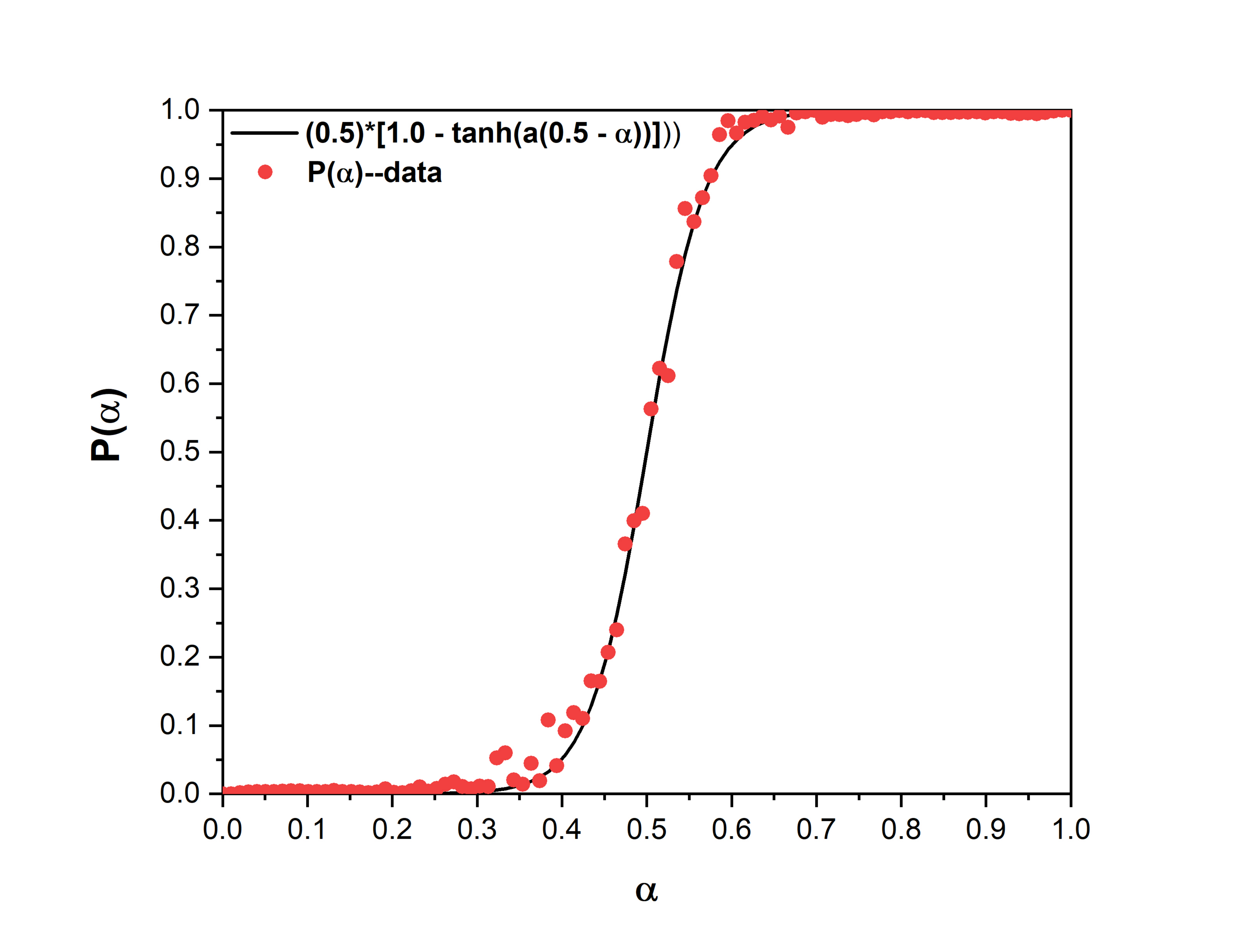}
    \caption{A WLC+SVM was trained on 50 baseline odors $I_k;\;k=1,2,\ldots,50$ and the mixture $\alpha I_1 + (1-\alpha)I_2$ presented.  Sigmoid curves were fit to the data.  The input mixture varies linearly in $\alpha$, while the classification probability follows a non-linear sigmoid.  The SVM enhances the separation in the AL and creates sharp boundaries between odors. The robustness to noise of the WLC+SVM network is due in part to the sharp boundary between classes of mixtures as shown here.}
    \label{fig: 2d_mix}
\end{figure}

Mixtures of odors are the biologically natural scenario for an insect exposed to the natural environment.  It is clearly evolutionary advantageous to distinguish between these separate odors in order to make decisions. We performed an experiment on the WLC+SVM network in which it is presented with mixtures of baseline currents Eq.\eqref{eq: input} ($\eta = 0$) and asked to classify the concentration of each. For the results displayed in Fig.(\ref{fig: 2d_mix}), the network was first trained on 50 such currents; from these, two were selected, and called $I_1$ and $I_2$. A mixture
\be
I_{mix}(\alpha) = \alpha I_1 + (1 - \alpha) I_2;\;\;\;0 \le \alpha \le 1
\ee
was then presented to the trained network, and the classification of $I_{mix}(\alpha)$ observed as a function of $\alpha$. $I_{mix}(\alpha)$ was classified as being either $I_1$ or $I_2$ with some very small leakage into the other 48 current classes. The classification is shown in Fig.(\ref{fig: 2d_mix}) and the results were expressed as probabilities using
\be
P_{I_k}(Current) = \frac{\mbox{Time in SVM region k}}{\mbox{Total time in all SVM regions}}.
\label{eq: prob}
\ee
Fig.(\ref{fig: 2d_mix}) shows $P_{I_1}(\alpha)$.  At $\alpha = 0$ it should be essentially zero, and at $\alpha = 1$, it should very close to unity. The fit for $P_{I_1}(\alpha)$ to the WLC+SVM voltage output data, shown in red dots, is  
\be
P_{I_1}(\alpha) = \frac{1}{2} \biggl [ 1 - \tanh{a(0.5 - \alpha)} \biggr ]\;\;= \frac{e^{a\alpha}}{e^{a\alpha} + e^{a(1-\alpha)}},
\label{eq: tanh}
\ee
with $a = 14.6$.

If the network is presented with a mixture which results in $P_{I_1}(\alpha)$, $\alpha$ (and $(1 - \alpha)$) can be recovered by inverting Eq.\eqref{eq: tanh}; thus the fraction of each pure odor is
\begin{equation*}
    \alpha = \frac{1}{2} + \frac{1}{a}\tanh^{-1}[2P_{I_1}(\alpha) - 1].
\end{equation*}

\subsection{Mixtures of Three or More Currents} \label{sec: mix3}

\begin{figure}[h!tbp] 
  \centering
  \includegraphics[width=4.17in,height=4.34in,keepaspectratio]{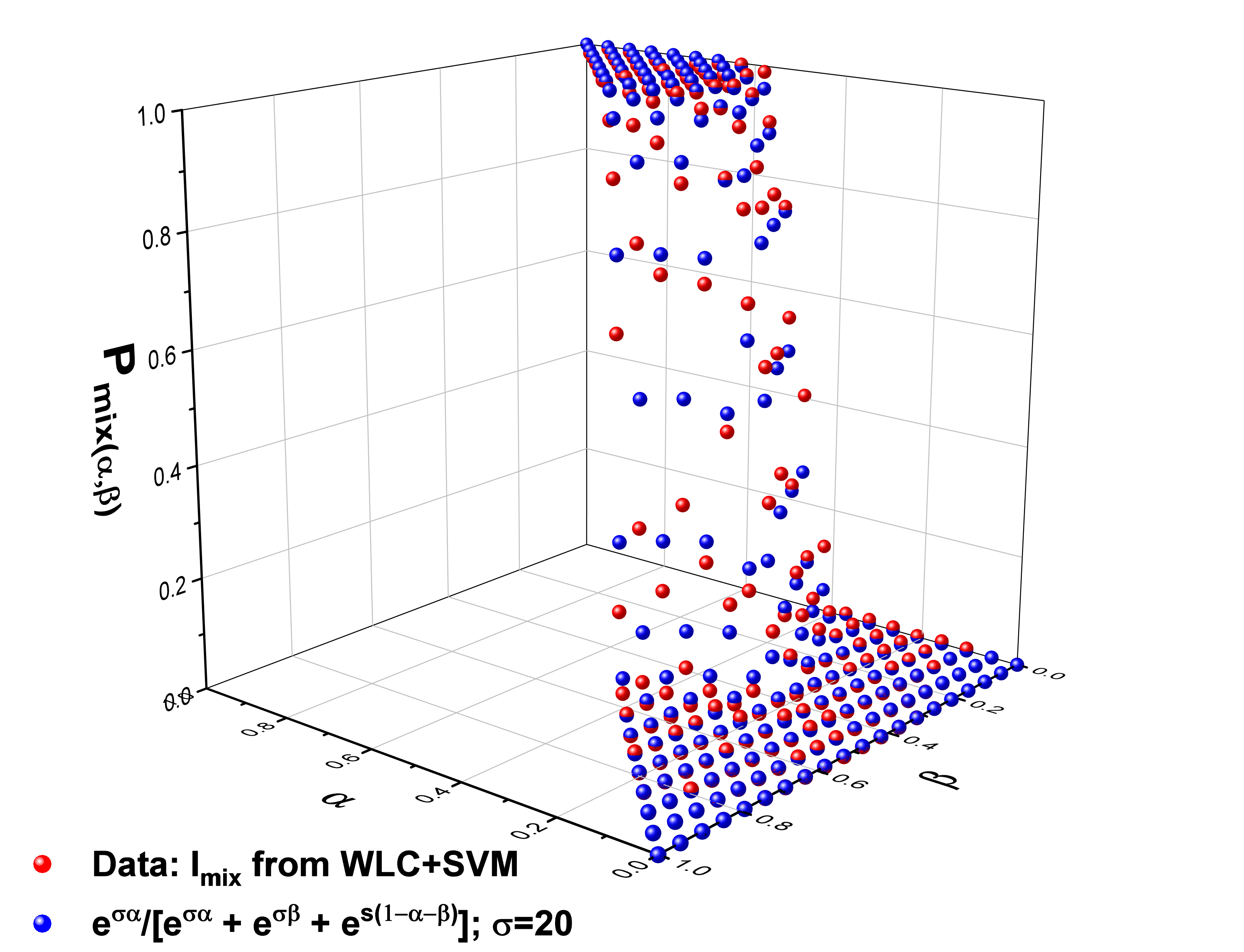}
  \caption{Mixture of three currents: $I_{mix}(\alpha,\beta) = \alpha I_1 + \beta I_2 +(1 - \alpha - \beta)I_3$ after processing by a 1000-dimensional WLC + SVM network (red dots).  Fit to data (blue dots) is given by
  $ P_1(\alpha,\beta) = e^{\sigma \alpha}/[e^{\sigma \alpha} + e^{\sigma \beta} + e^{\sigma(1- \alpha -\beta)}]$;
  $\sigma = 20.$}
  \label{fig: 3d_mix}
\end{figure}

The results of a mixture of two inputs in section \ref{sec: mix2} can be generalized to $K$ inputs in the following way. Given a linear mixture of $K$ base currents ($\balpha = \{\alpha_1,\alpha_2,\ldots,\alpha_K\};\sum_{k=1}^K \alpha_k = 1$)
\be 
\I_{mix, K}(\balpha) = \sum_{k = 1}^K\,\alpha_k \I_k,
\ee
the output from the WLC+SVM network is matched accurately by
\be
P_j(\balpha) = e^{\sigma \alpha_j}/\sum_{k=1}^K e^{\sigma \alpha_k};\;\;\sum_{k=1}^K \alpha_k = 1.
\ee
Inverting these equations gives the individual $\alpha_k$ as a function of the observed $P_j(\balpha,)$.  Fig.(\ref{fig: 3d_mix}) shows the three current mixture case with
\be
    I_{mix} = \alpha I_1 + \beta I_2 + (1 - \alpha-\beta)I_3
\ee
for $0 \le \alpha \le 1$, $0 \le\beta \le 1$, $\alpha + \beta \le 1$. 

\subsection{Time Dependent Stimulating Currents}\label{sec: time_dep}

All of the previous calculations were performed with step function input currents, but the usage of the WLC network is not limited to this particular class of inputs.  In this section we probe the network response to time dependent currents in the same fashion as before, by measuring the network activity over time and visualizing this activity in projected three dimensional PCA space.

The experiment was conducted by multiplying a time dependent signal $x_m(t);\;0 \le x_m(t) \le 1$ to the baseline inputs $\I_m$ Eq. \eqref{eq: input};  $\I_{m, \rm td} = x_m(t) *\I_m$.  All neurons in the WLC network receive the same current $x(t)$---which is rescaled from the $x$ variable of the Lorenz attractor \cite{lor63} to lie between 0 and 1---with parameters selected to make it chaotic.  Thus instead of a DC input set at $I_0$, the neurons received a time varying current varying between 0 and $I_0$.  The three tests for time dependent currents $\I_{1, \rm td}$, $\I_{2,\rm td}$, $\I_{3, \rm td}$ with temporal components $x_1$, $x_2$, $x_3$ and spatial components $\I_1$, $\I_2$, $\I_3$ are detailed in the table below:
\begin{center}
\begin{tabular}{ |c|c|c| } 
 test & spatial & temporal \\ 
 1 & $\I_1 \neq \I_2 \neq \I_3$  & $x_1 = x_2 = x_3$ \\ 
 2 &  $\I_1 \neq \I_2 \neq \I_3$  & $x_1 \neq x_2 \neq x_3$  \\ 
 3 &  $\I_1 = \I_2 = \I_3$  & $x_1 \neq x_2 \neq x_3$ \\
\end{tabular}
\end{center}

\begin{figure}[tbp]
  \includegraphics[width = 0.5\textwidth]{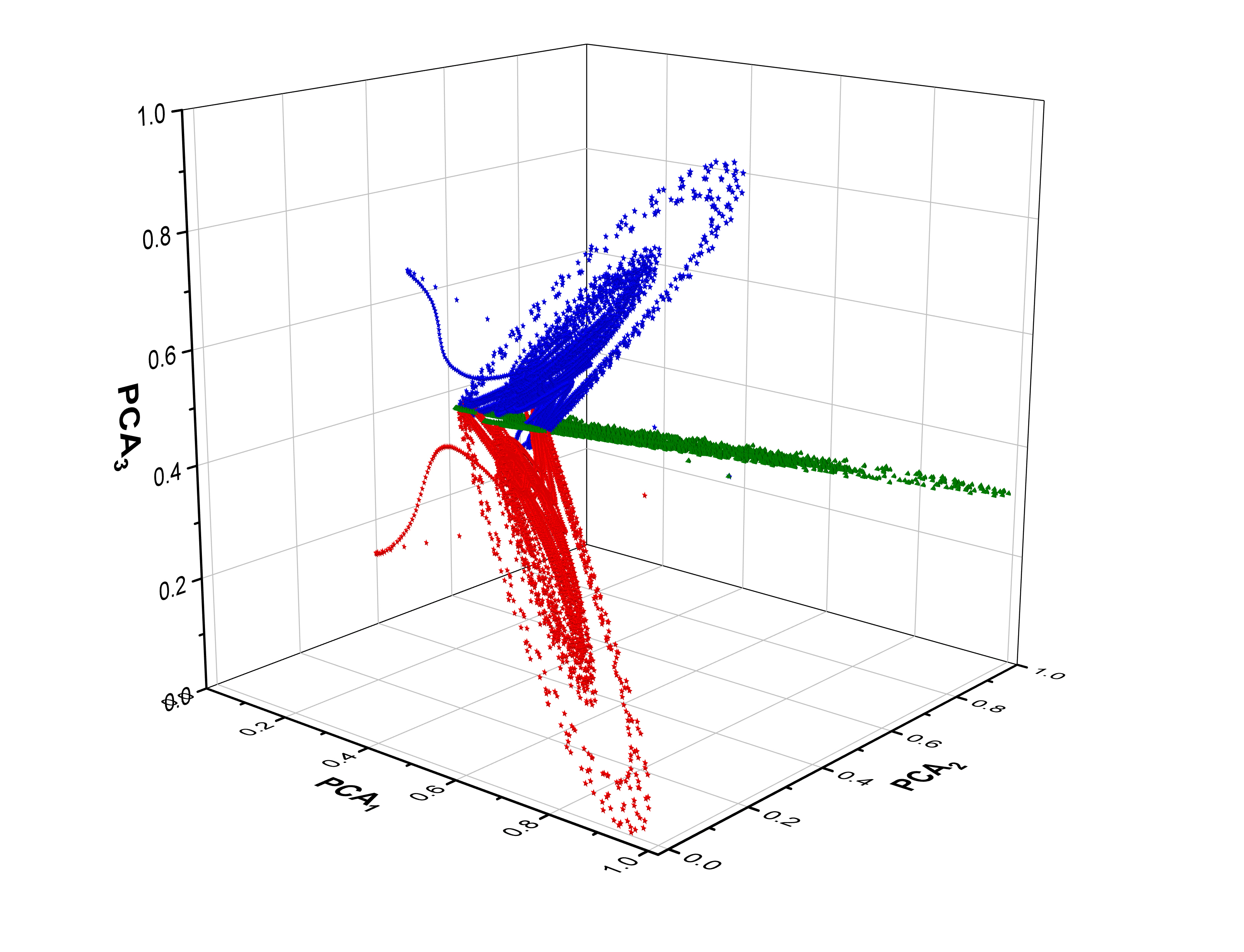}
    \includegraphics[width = 0.5\textwidth]{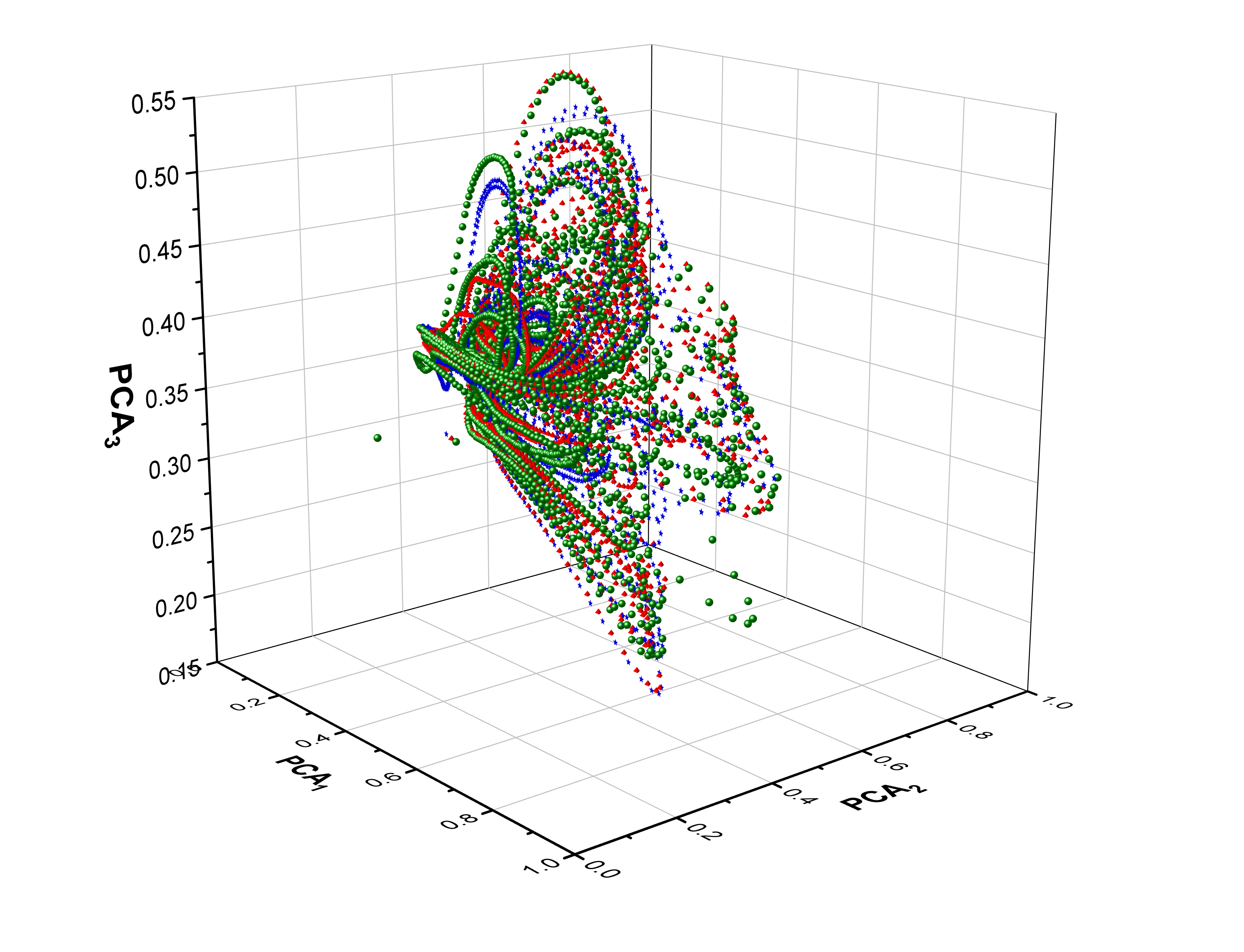}
  \caption{Activity of a $N$ = 1000 neuron WLC network built with FHN biophysical neurons Eq. (\ref{eq: FHN}) connected by inhibitory synaptic processes when stimulated by three distinct time dependent currents.  The currents in this case are constructed by taking a single chaotic time series and convolving this time series with three distinct base odors.  Thus the activity is the result of the same time series presented to three spatially distinct sets of neurons given by the baseline currents $\I_m$.  A PCA projection was performed as in Fig. (\ref{fig: pca_HH}). \textbf{(top)} \textbf{Test 1: } In this plot the time dependent amplitudes of the currents are all \textbf{identical}, with the difference only given by the spatial component \ie, the particular neurons that are injected.  Therefore, even though the time dependent currents are identical the WLC network separates them on the basis of their spatial composition.  This experiment is exactly analogous to the results of the previous section with the time dependent trajectory substituted for the basic step current. \textbf{(bottom)} \textbf{Test 3: } In this plot the time dependent amplitudes of the currents are all \textbf{different} but the spatial component is the same.  Therefore, the WLC network cannot separate out temporally distinct but spatially similar input currents in a way that could be classified by an SVM.}
\label{fig: time_dep}
\end{figure}

Test 1 involved different baseline currents with the same time series, test 2 different baseline currents and different time series, and test 3 the same baseline current and different time series. Fig. (\ref{fig: time_dep}) shows the results for tests 1 and 3 with the results from test 2 being qualitatively similar to test 1.  What Fig. (\ref{fig: time_dep}) tells us is that the WLC network is able to separate the activity of input currents if those currents are spatially separated, irrespective of the time dependence of the signal.  Both test 1 and test 2 show good separation in low dimensional PCA projections of the activity.  This characteristic is due to the fact that the spatial components of each of the input currents are distinct.  When we look at test 3 (corresponding to different time series but on the same neurons), however, there is not good spatial separation.  This lack of separation carries over to a SVM following the operation of the WLC network. Therefore temporally separated, but not spatially separated, signals will not be classified well by a WLC network followed by an SVM.

\subsection{SVM Alone}
\begin{figure}[h!tbp] % float placement: (h)ere, page (t)op, page (b)ottom, other (p)age
  \centering
  \includegraphics[width = 0.5\textwidth]{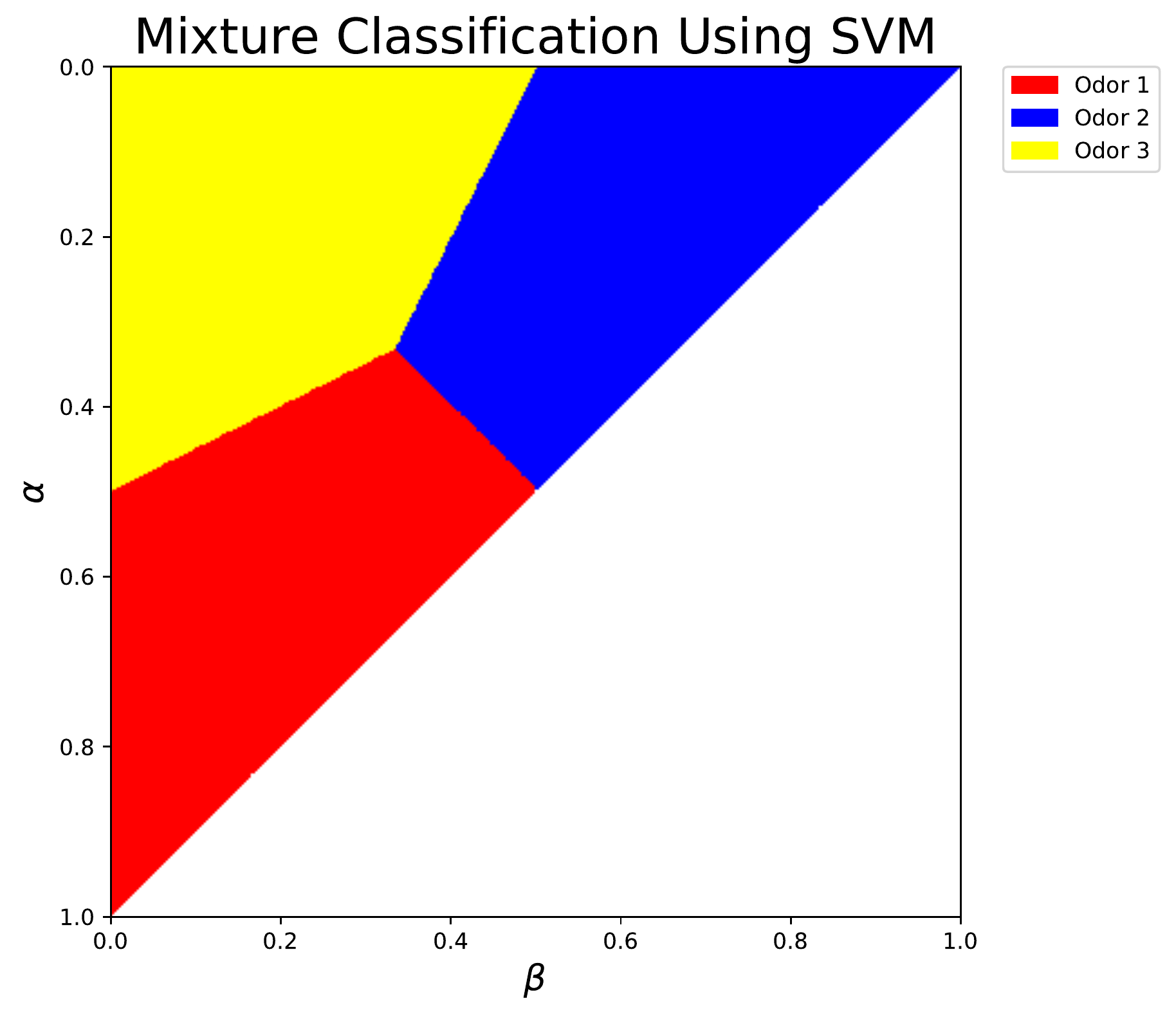}
  \caption{ A heat map display of the probability of a mixture of three currents using only an SVM and no WLC network. The sharp step function boundaries between the odor regions comes from the hyperplane boundaries given by the SVM.  This figure shows that the SVM alone cannot be used to evaluate the concentrations in a mixture $I_{mix}(\alpha, \beta) = \alpha I_1 + \beta I_2 + (1 - \alpha - \beta) I_3$.}
  \label{fig: svm_alone}
\end{figure}

An important question is why the WLC component appears to be used in animal olfaction. To examine this we removed the WLC, allowing an SVM to perform the classification alone on the input currents. Using an SVM is rather standard practice in ML classification networks. We found that removing the WLC from the network did not change the classification capability very much as we added noise $\eta > 0$ to the base currents $\I_m$, Eq.(\ref{eq: input}). However, {\bf the ability to separate mixtures of learned base currents was totally lost.}  Fig. (\ref{fig: svm_alone}) shows the decision probabilities for a mixture of three odors $I_{mix}(\alpha, \beta) = \alpha I_1 + \beta I_2 + (1 - \alpha - \beta) I_3$.  For an SVM alone we find that the discrimination that allowed calculation of $\alpha$ and $\beta$ for the WLC+SVM network disappears.

\section{Conclusion and Discussion}
The input stimulus points in a direction in N-dimensional space and generates an orbit defining the attractor, which the network activity traverses as long as the stimulus is on. Even when perturbed by high noise levels or mixed with other odors, this ``net'' directionality keeps the activity of the network localized. Perturbations \ie, $\eta > 0$ Eq.(\ref{eq: input}), disturb both the magnitude of the input base currents and their directions. The SVM  interprets the localized activity as being indicative of a particular, learned current and then optimally separates them. If one takes a mainstream ML approach by eliminating the WLC part of the olfaction informed network, there is a total reduction in the capability to separate odor (current) mixtures.

An item not studied in depth, but requiring more interpretation, is the exact trajectory of the network around the WLC attractor. The theoretical capacity of the network~\cite{rabinovich2001} may be fully realized only when each trajectory can be identified as associated with its own current.  As seen in section \ref{sec: time_dep}, the WLC network is unable to separate out currents with different time varying inputs to the same set of neurons.  The SVM amalgamates similar inputs in the same region of space into a single current, masking the temporal information in the signal.  
%This particular question of chaotic time series prediction and classification suggests an interesting connection with other work done by Ott, Hunt and colleagues~\cite{ott18,ottdresden19} where a fixed, randomly connected, high dimensional `reservoir' network with a linear output layer is trained to recognize patterns in both low and high dimensional nonlinear systems. Only the output layer is trained. The connection to this work may be that the reservoir can act as a WLC network---the universality of large random networks with inhibitory connections operating as WLC devices is conjectured in~\cite{rabinovich2001}. 

We call attention to another investigation of an insect olfactory system and its ability to classify~\cite{delahunt_biological_2018}. The AL in that work was represented as a noisy relaxation oscillator. The AL presented here is a richer system with significant capacity to classify. It does not act as a noisy circuit, but a deterministic chaotic device~\cite{rabinovich2001}.

The results of this paper on robustness against noise and reliability in identification of mixtures of objects in a class---using a WLC+SVM network abstracted from a functional biological neural device---removes some of the `mystery' of how ML networks operate. Using this framework it is possible to understand the physical mechanisms behind the success of this kind of supervised learning network. This WLC+SVM network realized in biomimetic circuitry need not operate at about 10Hz, which it does when constructed from biological material in locusts.

\section{Acknowledgments}
We acknowledge partial support from Microsoft Research. Work done by David Li and Jonathan Lam helped frame the questions in this paper. Conversations with Charles Delahunt of the University of Washington and Mark A. Stopfer of the National Institutes of Health on the insect olfactory system have been useful in the development of this work.  We acknowledge helpful discussions with Maxim Bazhenov on models of the antennal lobe.

\newpage
\bibliography{prr}
\end{document}